\documentclass[twocolumn,showpacs,amsmath,amssymb]{revtex4}
\pdfoutput=1

\usepackage{graphicx}
\usepackage{dcolumn}
\usepackage{bm}
\newcommand{\etal}{\emph{et al.}}
\newcommand{\be}{\begin{equation}}
\newcommand{\ee}{\end{equation}}
\newcommand{\bfig}{\begin{figure}}
\newcommand{\efig}{\end{figure}}

\begin{document}      
\title{Anomalous conductivity tensor in the Dirac semimetal $\mathbf{Na_3Bi}$
} 
 
\author{Jun Xiong$^{1}$}
\author{Satya Kushwaha$^{2}$}
\author{Jason Krizan$^2$}
\author{Tian Liang$^1$}
\author{R. J. Cava$^2$}
\author{N. P. Ong$^{1}$}
\affiliation{
Departments of Physics$^1$ and Chemistry$^2$, Princeton University, Princeton, NJ 08544
} 

\date{\today}      
\begin{abstract}
Na$_3$Bi  is a Dirac semimetal with protected nodes that may be sensitive to the breaking of time-reversal invariance in a magnetic field $\bf B$. We report experiments which reveal that both the conductivity and resistivity tensors exhibit robust anomalies in $B$. The resistivity $\rho_{xx}$ is $B$-linear up to 35 T, while the Hall angle exhibits an unusual profile approaching a step-function. The conductivities $\sigma_{xx}$ and $\sigma_{xy}$ share identical power-law dependences at large $B$. We propose that these significant deviations from conventional transport result from an unusual sensitivity of the transport lifetime to $B$. Comparison with Cd$_3$As$_2$ is made. 
\end{abstract}
 
\maketitle      
In Dirac semimetals -- the analogs of graphene in three-dimensional material -- the bulk Dirac node is protected against gap-formation via hybridization. The iridate pyrochlores were initially predicted~\cite{Vishwanath} to have protected nodes, but crystal growth has been problematical. Recently, Young \etal~\cite{Young} identified a class of materials in which time-reversal invariance (TRI) leads to node protection when the nodes occur at high-symmetry points (the time-reversal invariant momenta or TRIM). Subsequently, Wang et al. proposed that crystalline symmetry can protect Dirac nodes even when they occur away from TRIM. From band calculations, they identified Na$_3$Bi~\cite{Wang1} and Cd$_3$As$_2$~\cite{Wang2} as Dirac semimetals. Photoemission~\cite{Chen,Neupane,Liu,Borisenko,Suyang} and scanning tunneling microscopy~\cite{Yazdani} recently confirmed that bulk Dirac nodes exist in both Na$_3$Bi and Cd$_3$As$_2$. Many groups~\cite{Son,Balents,Hosur,Sid} predict that, in Dirac and Weyl semimetals, charge pumping associated with the chiral anomaly can be observed in an intense magnetic field $\bf B$. We report experiments on Na$_3$Bi showing that, even at low $B$, the breaking of TRI leads to robust anomalies in the conductivity tensor which appears to originate from a strongly $B$-dependent transport lifetime $\tau_{tr}(B)$. Among the anomalies are a robust $B$-linear magnetoresistence and an unusual step-like field profile of the Hall angle $\tan\theta$. The quantum oscillations show that Fermi surface (FS) is comprised of two distinct valleys, i.e. the Fermi energy lies in the non-trivial gap-inverted regime.

\begin{figure*}[h]
\includegraphics[width=14 cm]{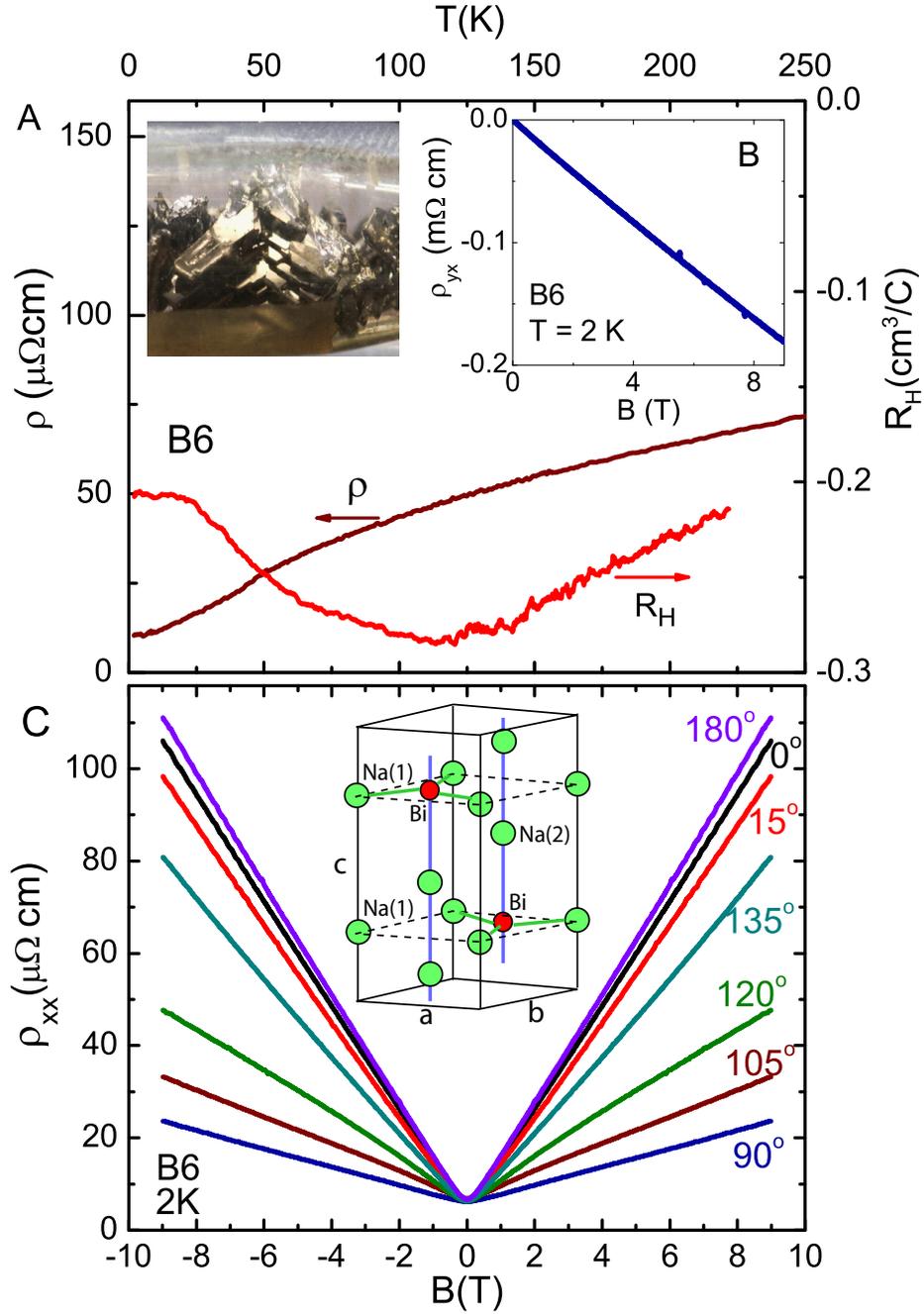}
\caption{\label{figR} (Color online)
Magnetotransport in Na$_3$Bi. Panel A: The zero-field resistivity $\rho$ and the Hall coefficient $R_H$ vs. $T$ (measured with $\bf H||c$). Inset shows the crystals sealed in a vial. The largest facet is normal to $\bf\hat{c}$. Panel (B) shows the Hall resistivity $\rho_{yx}$ vs. $B$ measured at 2 K in B6.
Panel (C): The $H$-linear magnetoresistance in Sample B6 measured at 2 K at selected tilt angles $\theta$ to $\bf\hat{c}$. The MR ratio is largest at $\theta=0^\circ$ (and 180$^\circ$). (${\bf B} = \mu_0 {\bf H}$ with $\mu_0$ the vacuum permeability). The crystal structure of Na$_3$Bi is sketched in the inset (adapted from Ref.~\cite{Chen}). 
}
\end{figure*}

Na$_3$Bi single crystals were crystallized from the Na-rich compositions (90 and 95$\%$) tuned to preclude the formation of the superconductor NaBi as an impurity phase~\cite{Satya,Brauer} (growth details are published elsewhere). The crystal structure was confirmed by X-ray diffraction (the lattice structure is sketched in Fig. \ref{figR}C). The deep-purple crystals grow with the largest facets normal to the $c$-axis (001) . To avoid deterioration of the crystals (which fully oxidize within 30 s of exposure to air), we attached contacts with Ag epoxy to the crystals inside an Ar glove-box and then covered them with oil before transferring to the cryostat (see Supplement). The resistivity profile $\rho$ vs. $T$ is metallic (Fig. \ref{figR}A) with residual values ranging from 1.72 to 87 $\mu\Omega$cm. The Hall resistivity $\rho_{yx}$ is $n$-type and strictly $B$-linear (Fig. \ref{figR}), with a nearly $T$-independent Hall coefficient $R_H = \rho_{yx}/B$ ($\bf\hat{x}||I$ and $\bf\hat{z}||\hat{c}$, where $\bf I$ is the current). Batch B and C samples were measured without post annealing, while G1 was post-annealed for 1 month. Table S1 in the Supplement lists the transport quantities measured in 8 samples.

Figure \ref{figR}B plots the MR curves in Sample B6 for selected $\theta$ (the tilt angle between $\bf\hat{c}$ and the field ${\bf H}={\bf B}/\mu_0$, with $\mu_0$ the vacuum permeability). As shown by the fan pattern, the MR ratio $\rho_{xx}(B)/\rho_{xx}(0)$ decreases rapidly as $\bf H$ is tilted into the $a$-$b$ plane ($\theta\to 90^\circ$). 

\begin{figure*}[t]
\includegraphics[width=14 cm]{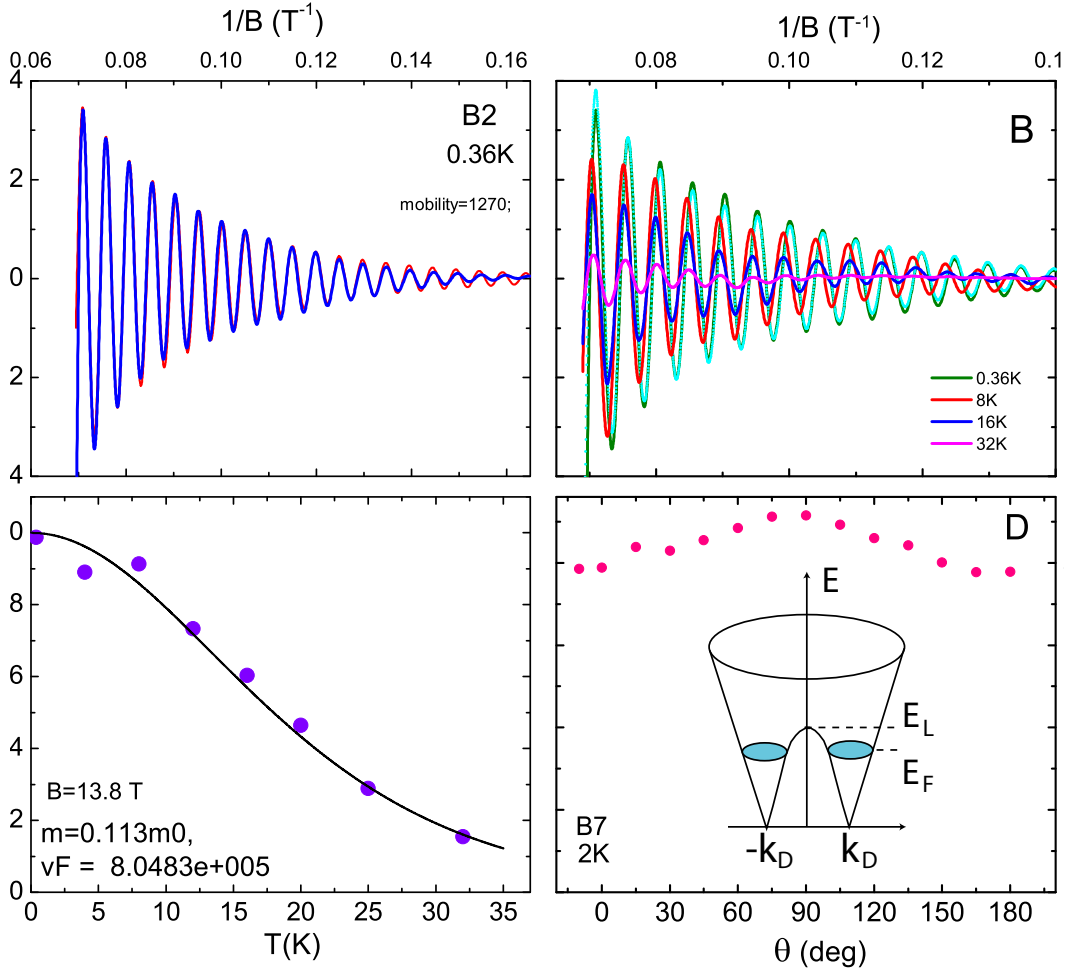}
\caption{\label{figtorque} (Color online)
Torque measurements of the de-Haas van Alven (dHvA) oscillations in Na$_3$Bi. The dHvA oscillations (solid curve in Panel A) can be fit well to the LK expression (dashed curve) with one period. From the damping versus $H$ (Panel A) and the $T$ dependence (Panels B and C), we obtain the Fermi surface section $S_F$, the effective mass $m^* = 0.11 m_0$ ($m_0$ the free mass), velocity $v_F$ = 8.05$\times 10^5$ m/s, and $\tau_Q$ = 8.16$\times 10^{-14}$s. The plot of the peak fields $B_{min}$ and $B_{max}$ vs. the integers $N$ (Panel D) yields $S_F$. 
}
\end{figure*}

All the samples display prominent Shubnikov de Haas (SdH) oscillations in $\rho_{xx}(B)$, from which the Fermi Surface cross-section $S_F$ and the Fermi wavevector $k_F$ are determined. In addition, we have measured the de Haas van Alven (dHvA) oscillations using torque magnetometry. Figure \ref{figtorque}A shows the oscillatory component of the magnetization together with a fit to the Lifshitz-Kosevich (LK) expression. From fits to the dHvA amplitudes vs. $1/H$ and $T$ (Panels B and C), we determine the effective mass $m^*$, the Fermi velocity $v_F$ and the quantum lifetime $\tau_Q$. The carrier density $n = g_vk_F^3/3\pi^2$ (with the valley and spin degeneracies $g_v$ and $g_s$ both equal to 2) ranges from 2.6-4.1$\times 10^{19}$ cm$^{-3}$, consistent with the Hall effect (Table S1). In addition, by varying $\theta$ in the MR, we verify that the FS cross section $S_F$ is nearly spherical (Fig. \ref{figtorque}D).

\begin{figure*}[t]
\includegraphics[width=14 cm]{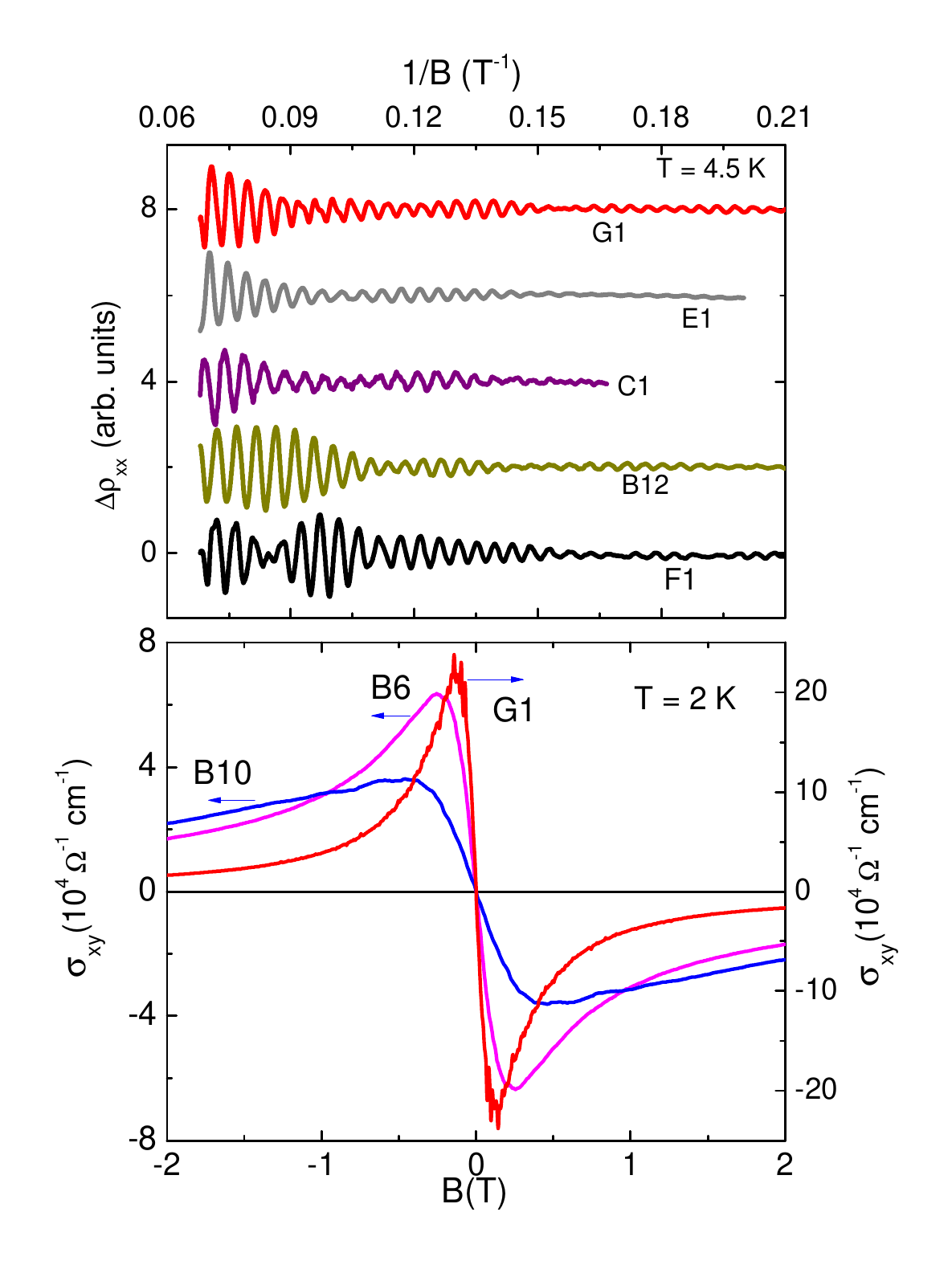}
\caption{\label{figBeats} (Color online)
Panel A: Curves of $\Delta\rho_{xx}$ showing modulation of the SdH amplitude. The beating pattern implies a small splitting of the fundamental SdH frequency.
Panel B compares the Hall conductivity $\sigma_{xy}(B)$ in 3 samples at 2 K. In the 3 curves, the peak field $\pm B_{\mu}$ yields the values $\mu$ = 21,640, 39,250 and 91,000 cm$^2$/Vs in B10, B6 and G1, respectively.
}
\end{figure*}

Because of unintentional doping from vacancies, the Fermi energy $E_F$ is in the conduction band, as implied by the $n$-type sign of the Hall resistivity $\rho_{yx}$. The band calculations~\cite{Wang1} predicted the existence of two Dirac nodes centered at $(0,0,\pm k_D)$ caused by gap inversion (sketch in Fig. \ref{figtorque}D). As $E_F$ rises in the conduction band, the two Dirac cones merge into a single cone when $E_F$ exceeds the Lifshitz-transition energy $E_L$. Recent ARPES experiments have confirmed the predicted dispersion and measured $k_D$ to be 0.095 $\rm{\AA}^{-1}$~\cite{Chen} and 0.10 $\rm{\AA}^{-1}$~\cite{Suyang} (see also~\cite{ZXShen}). An important issue is whether $E_F$ in our samples lies above or below $E_L$. As shown in Table S1, our measured $k_F$ ranges from 0.073 to 0.084 $\rm{\AA}^{-1}$. As these values are smaller than $k_D$, we conclude that $E_F$ lies below $E_L$. The FS is comprised of two distinct valleys on opposite sides of $\Gamma$, i.e. $g_v = 2$. A persistent feature is a weak beating pattern in the SdH oscillations. Figure \ref{figBeats}a shows the SdH traces from 5 samples. The Fourier spectra of the oscillations reveal two frequencies $f_1$ and $f_2$ corresponding to two values of $S_F$ differing by $\sim 16\%$. The beating suggests that the orbits may reflect quantum interference between the two orbits. We hope to explore the beating in high-mobility crystals.

The mobility $\mu$ of each sample is directly measured from the field profile of the Hall conductivity $\sigma_{xy}$. As shown in Fig. \ref{figBeats}b, $\sigma_{xy}(B)$ has the characteristic dispersion-resonance profile produced by cyclotron motion of the carriers. By the Bloch-Boltzmann theory, the extrema in $\sigma_{xy}(B)$ occur at the peak fields $\pm B_\mu$, with $1/B_\mu = \mu$. With increasing $\mu$, from sample B10 ($\mu$ = 21,640 cm$^2$/Vs) to B6 (39,250 cm$^2$/Vs) and G1 (91,000 cm$^2$/Vs), the peak field $B_{\mu}$ systematically decreases. The variation in $\mu$ strongly influences the Hall-angle profile (see below). Using $\mu = ev_F\tau_{tr}/\hbar k_F$, we find that the transport lifetime $\tau_{tr}$ exceeds $\tau_Q$ by a ratio $R_{\tau}$ = 10-20 ($R_{\tau}$ is expected to exceed 1 since $\tau_Q$ reflects broadening due to all scattering processes). In Cd$_3$As$_2$, $R_{\tau}$ is as large as 10$^4$~\cite{Liang}.

In the relaxation-time $\emph{ansatz}$, the Boltzmann equation describing changes to the distribution function $f_{\bf k}$ caused by an electric field $\bf E$ is expressed as~\cite{Ziman}
\be
e{\bf E\cdot v}\frac{\partial f^0_{\bf k}}{\partial E_{\bf k}}  
+ e{\bf v\times B\cdot}\frac{\partial g_{\bf k}}{\partial {\bf k}}
= -\frac{g_{\bf k}}{\tau_{tr}},
\label{Boltz}
\ee
where $e$ is the elemental charge and $g_{\bf k} = f_{\bf k}-f^0_{\bf k}$, with $f^0_{\bf k}$ the Fermi-Dirac function. $E_{\bf k}$ and velocity $\bf v$ are, respectively, the energy and velocity at state $\bf k$. The $\emph{ansatz}$ yields the conductivity tensor $\sigma_{ij}$, with 
\be
\sigma_{xx} = ne\mu/D, \quad\quad \sigma_{xy} = ne\mu^2B/D,
\label{sxx}
\ee
where $D = 1+(\mu B)^2$. From Eq. \ref{sxx}, the ratio $\sigma_{xy}/\sigma_{xx}= \mu B$, which is the Hall angle $\tan\theta$, is linear in $B$. By contrast, the resistivity $\rho_{xx}$ is $B$-independent because the Hall electric-field $E_y$ exactly balances the Lorentz force. Significantly, $\sigma_{xx}\sim 1/B^2$ decreases much faster than $\sigma_{xy}\sim 1/B$ when $\mu B\gg 1$. These standard predictions assume that $\tau_{tr}$ (hence $\mu$) is a constant independent of $B$. In conventional metals and semimetals in the impurity-scattering regime (elastic scattering), this assumption is firmly established; the predicted trends are a cornerstone of semiclassical transport.

\begin{figure*}[t]
\includegraphics[width=14 cm]{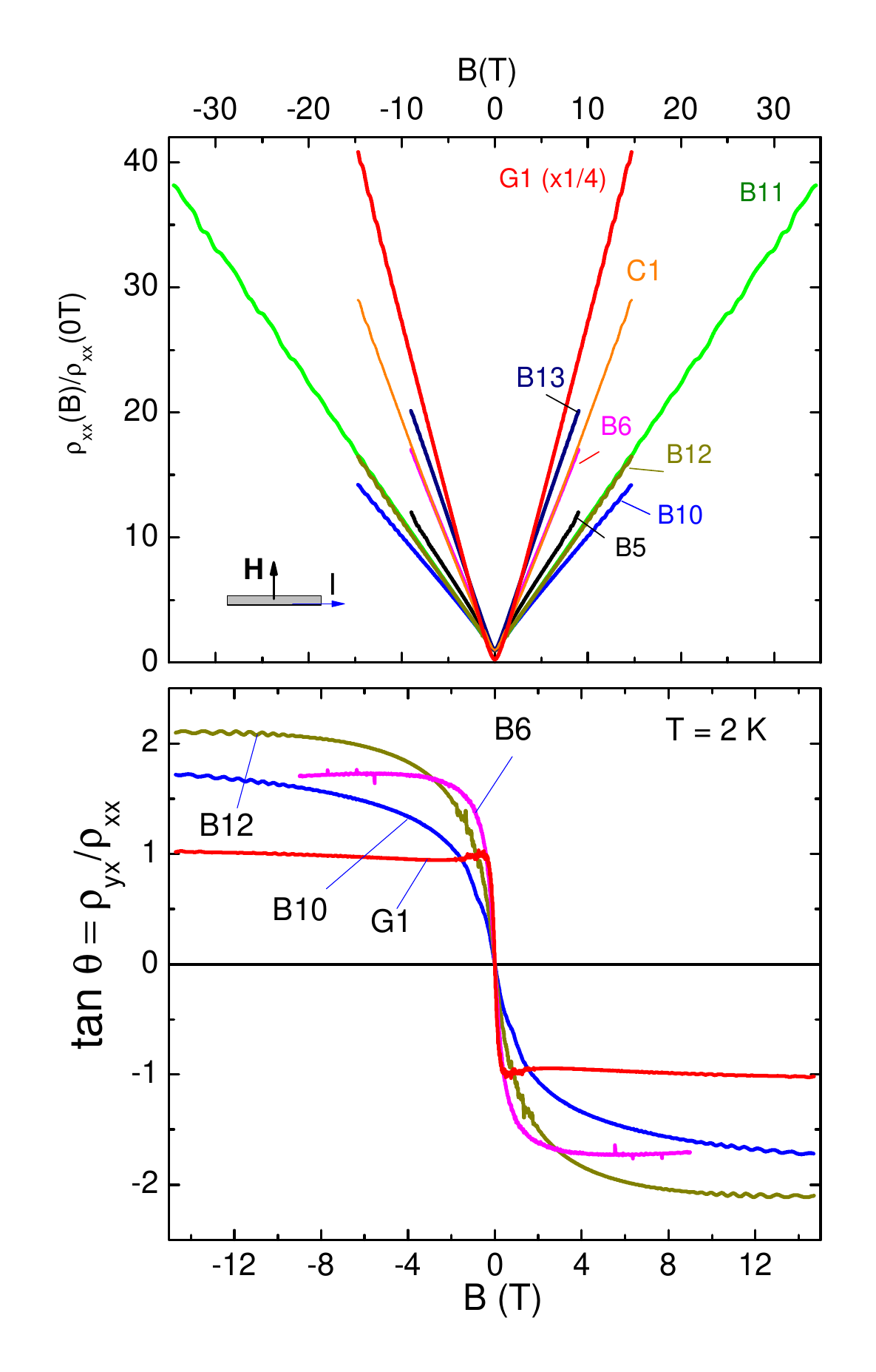}
\caption{\label{figMRtan} (Color online)
Robust $H$-linear magnetoresistance in Na$_3$Bi (Panel A). In the 8 samples shown, $\rho_{xx}(B)$ is measured with $\bf H||\hat{c}$ at 2 K in all cases except in B11 (at 1.6 K). In B11, the MR persists without observable deviation to 35 T. A general trend is that the MR increases with $\mu$, from $\sim 14$ (at 15 T) in B10, to $\sim 160$ in G1 (which has the highest $\mu$) (the MR in G1 is plotted in $\frac14$ scale). In Panel B, the field profile of $\tan\theta = \rho_{yx}/\rho_{xx}$ is compared in 4 samples. As $H$ increases, $\tan\theta$ rapidly saturates to an $H$-independent value, which implies the anomalous relationship $\tau_{tr}\sim 1/H$. In G1, the change occurs at 0.5 T.}
\end{figure*}

In Na$_3$Bi, however, the observed field dependences of the diagonal elements $\sigma_{xx}$ and $\rho_{xx}$ disagree in an essential way from the standard predictions (only $\rho_{yx}(B)$ appears conventional). As we noted in Fig. \ref{figR}, the resistivity increases linearly with $B$ instead of saturating. A $B$-linear MR is rare in conventional conductors (see Abrikosov's comments~\cite{Abrikosov}; we exclude metals with open orbits~\cite{Ziman}). However, in materials with unusual topological phases, a $B$-linear MR is increasingly encountered~\cite{Qu}. To persuade ourselves that the $B$-linear MR is pervasive in Na$_3$Bi, we have investigated 8 samples (Table S1). Figure \ref{figMRtan}A shows that the $B$-linear MR is a very robust feature in Na$_3$Bi. Across the samples, the MR ratio (measured at 15 T) increases from $\sim 14$ in B10, to $163$ in G1 (the sample with the highest $\mu$). In B11, we show that the $B$-linear profile extends to 35 T with no evidence of deviation.

A second dramatic anomaly is seen in the Hall-angle profile. 
In Fig. \ref{figMRtan}B, we compare $\tan\theta$ measured in four samples with increasing $\mu$, B10, B12, B6 and G1. As shown, $\tan\theta$ initially rises very rapidly in weak $B$ at a rate dictated by the mobility, but saturates to a plateau value at large $B$. Whereas the saturation is gradual in the samples with low mobility (B10 and B12), the rise becomes abrupt in higher-mobility samples (B6 and G1). In G1, especially, the profile resembles a step-function profile. $\tan\theta$ assumes a virtually $B$-independent value from $B$ = 0.5 to 15 T instead of increasing linearly with $B$. Since $\tan\theta = \mu B$, the simplest interpretation of the step-function profile is that, starting in weak $B$, the transport lifetime varies with $B$ as 
\be
\tau_{tr} \sim 1/B.
\label{tau}
\ee

The merit of Eq. \ref{tau} is that it also accounts for the $B$-linear MR profile, i.e. 
$\rho_{xx} = 1/ne\tau_{tr} \sim B.$
Interestingly, with Eq. \ref{tau}, the high-field $B$ dependence of $\sigma_{xx}$ is reduced by one power of $B$ to $\sigma_{xx}(B)\sim 1/B$ (Eq. \ref{sxx}), but leaves that of $\sigma_{xy}\sim 1/B$ unchanged because $\mu$ cancels out at large $B$. Hence both $\sigma_{xx}$ and $\sigma_{xy}$ vary as $1/B$ at large $B$, consistent with the step-profile of $\tan\theta$. 

\begin{figure*}[t]
\includegraphics[width=14 cm]{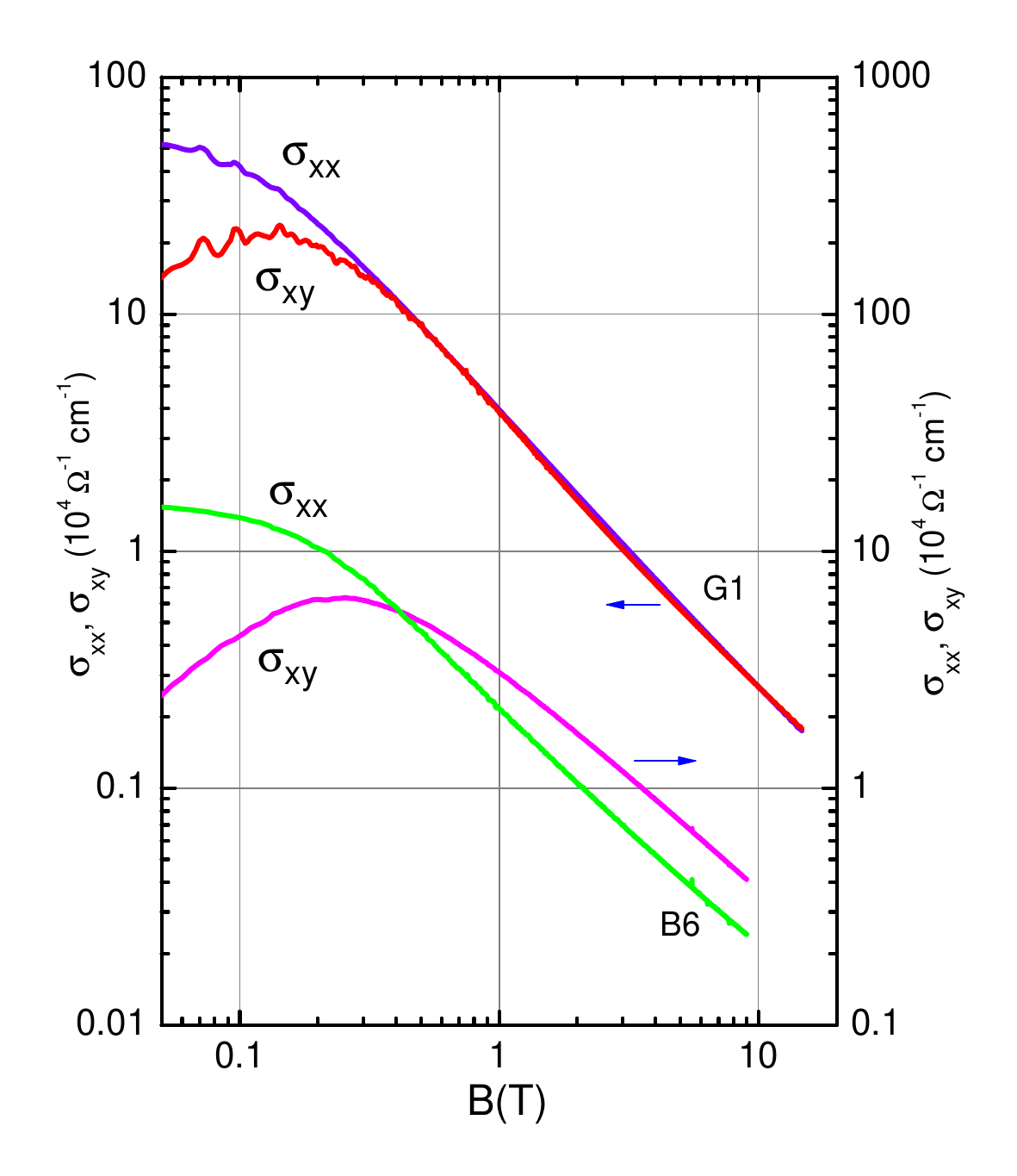}
\caption{\label{figCond} (Color online)
Log-log plots of $\sigma_{xx}$ and $\sigma_{xy}$ vs. $B$ in G1 and B6. Consistent with Eq. \ref{tau}, both quantities approach the same power law $B^{-\beta}$ when $B$ exceeds 0.3 and 2 T in G1 and B6, respectively. The measured $\beta$ is 1.15 and 1.0 in G1 and B6, respectively. Curves for B6 are shifted vertically.
}
\end{figure*}

To verify this, we plot the $B$ dependences of $\sigma_{xx}$ and $\sigma_{xy}$ in log-log scale for the two high-mobility samples B6 and G1 (Fig. \ref{figCond}). In both samples, the two conductivities have the same power-law dependence $B^{-\beta}$ above a relatively low $B$. Consistent with the behavior of $\tan\theta$, this occurs at $B$ = 2 T and 0.3 T in B6 and G1, respectively. The measured value of $\beta$ is 1.0 in B6, but is slightly larger (1.15) in G1. 

Theoretically, in the Weyl semimetal, the Weyl nodes which come in pairs act as sources and sinks of Berry curvature (Chern flux)~\cite{Vishwanath,Son,Balents,Hosur,Sid}. To realize a finite Berry curvature $\vec{\Omega}(\bf k)$, TRI must be broken in the Weyl semimetal. The 3D Dirac semimetal may be regarded as the limiting case when TRI is restored. In this limit, the Weyl nodes coincide in $\bf k$ space but are prevented from hybridizing by crystalline symmetry. Conversely~\cite{Wang1}, one expects that, in a Dirac semimetal, the breaking of TRI by an applied $\bf B$ will render $\vec{\Omega}(\bf k)$ finite, and lead to essential changes in its FS best detected by transport experiments. 

To date, theoretical attention has focused on the charge-pumping of the chiral current from left to right-moving branches~\cite{Son,Balents,Hosur,Sid,Nielsen}. However, as shown by the results here, the essential changes to the FS topology induced by $\bf B$ should skew $\sigma_{xx}$ and $\rho_{xx}$ away from conventional behavior even in weak $B$. The results for Na$_3$Bi (and Cd$_3$As$_2$~\cite{Liang}) invite a systematic re-examination of transport in Dirac semimetals. We compare the two materials in the Supplement.


\centerline{*    *     *}\vspace{6in}
{We thank Andrei Bernevig and Ashvin Vishwanath for valuable discussions. N.P.O. is supported by 
the Army Research Office (ARO W911NF-11-1-0379). R.J.C. and N.P.O. are supported by a MURI grant on Topological Insulators (ARO W911NF-12-1-0461) and by the US National Science Foundation
(grant number DMR 1420541). T.L. acknowledges scholarship support from the Japan Student Services Organization. Some of the experiments were performed at the National High Magnetic Field Laboratory, which is supported by National Science Foundation Cooperative Agreement No. DMR-1157490, the State of Florida, and the U.S. Department of Energy.}

\vspace{3cm}

\newpage

\end{document}